\documentclass{aa}
\usepackage{subfig}
\usepackage{amsmath}
\usepackage{times}
\usepackage{natbib}
\usepackage[colorlinks,linkcolor=blue,anchorcolor=blue,citecolor=blue]{hyperref}
\usepackage[hyperref]{backref}
\usepackage{txfonts}
\usepackage[T1]{fontenc} 

\usepackage{graphicx}
\usepackage{color}
\usepackage{natbib}
\usepackage{rotating}
\usepackage{longtable,lscape}

\begin{document}
   \title{\textit{Chandra} and optical/IR observations of CXO
     J1415.2+3610, a massive, newly discovered galaxy cluster at
     $z\sim 1.5$}

\author{P. Tozzi \inst{1,2}, J.S. Santos \inst{3}, M. Nonino
  \inst{1}, P. Rosati \inst{4}, S. Borgani \inst{1,2,5}, B. Sartoris
  \inst{6,7}, B. Altieri \inst{3} and M. Sanchez-Portal \inst{3} }

\institute{ \inst{1} INAF-Osservatorio Astronomico di Trieste, Via
  Tiepolo 11, 34133 Trieste, Italia \\ \inst{2} INFN-National
  Institute for Nuclear Physics, via Valerio 2, I--34127, Trieste
  \\ \email{tozzi@oats.inaf.it} \\ \inst{3} European Space Astronomy
  Centre (ESAC)/ESA, Madrid, Spain \\ \inst{4} European Southern
  Observatory, Karl-Schwarzschild Strasse 2, 85748 Garching, Germany
  \\ \inst{5} Universit\`a di Trieste, Dipartimento di Fisica, via
  Valerio, 2 - 34127 Trieste \\ \inst{6} University Observatory,
  Ludwig-Maximillians University Munich, Scheinerstrasse 1, 81679
  Munich, Germany \\ \inst{7} Excellence Cluster Universe,
  Boltzmannstrasse 2, 85748 Garching, Germany }

   \date{Received ... ; accepted ...}

 
 \abstract
   {} 
{We report the discovery of CXO J1415.2+3610, a distant ($z\sim$1.5)
  galaxy cluster serendipitously detected as an extended source with a
  very high significance level ($S/N \sim 11$) in a deep,
  high-resolution \textit{Chandra} observation targeted to study the
  cluster WARP J1415.1+3612 at $z=1.03$.  This is the highest-z
  cluster discovered with \textit{Chandra} so far.  Moreover, the total
  exposure time of 280 ks with ACIS-S provides the deepest X-ray
  observation currently achieved on a cluster at $z\geq 1.5$.}
{We perform an X-ray spectral fit of the extended emission of
  the intracluster medium (ICM) with \textit{XSPEC} assuming a
  single-temperature thermal \textit{mekal} model.  We use optical and infrared (IR)
  observations from Subaru-\textit{Suprime} (BVRiz), \textit{MOIRCS}
  (JKs), and \textit{Spitzer}/IRAC (3.6 $\mu$m) to confirm the
  presence of an overdensity of red galaxies matching the X-ray
  extended emission.  We use optical and IR data to investigate the
  color-magnitude relation of the candidate member galaxies.  }
{From a preliminary X-ray spectral analysis, we detect at a $99.5$\%
  confidence level the rest frame 6.7-6.9 keV Iron $K_\alpha$ line
  complex, from which we obtain $z_X = 1.46\pm 0.025$.  Our X-ray
  redshift measurement is supported by the optical and IR data.  The
  analysis of the $z-3.6\mu$m color-magnitude diagram shows a well-defined 
  sequence of red galaxies within $1\arcmin$ from the cluster
  X-ray emission peak with a color range [$5 < z-3.6 \mu m < 6$].
  The photometric redshift obtained by spectral energy distribution (SED) fitting is $z_{phot} =1.52
  \pm 0.06$.  After fixing the redshift to $z=1.46$, we perform the
  final spectral analysis and measure the average gas temperature with
  a 20\% error, $kT=5.8^{+1.2}_{-1.0}$ keV, and the Fe abundance
  $Z_{Fe} = 1.3_{-0.5}^{+0.8}Z_\odot$.  We fit the background-subtracted 
  surface brightness with a single beta-model out to 35
  arcsec (the maximum radius where the X-ray emission is detected),
  and derive the deprojected electron density profile.  The ICM mass
  is $1.09_{-0.2}^{+0.3}\times 10^{13} M_\odot$ within $300$ kpc.
  Under the assumption of hydrostatic equilibrium, the total mass is
  $M_{2500}= 8.6_{-1.7}^{+2.1} \times 10 ^{13} M_\odot$ for $R_{2500}
  = (220\pm 55)$ kpc.  Extrapolating the profile at larger radii, we
  find $M_{500}= 2.1_{-0.5}^{+0.7} \times 10 ^{14} M_\odot$ for
  $R_{500} = 510_{-50}^{+55}$ kpc.  This analysis establishes
  CXOJ1415.2+3610 as one of the best characterized distant galaxy
  clusters based on X-ray data alone. }
 {}


\authorrunning{P. Tozzi et al.}  

\titlerunning{A massive, newly discovered cluster at $z\sim 1.5$}

\maketitle
  

\section{Introduction}

Clusters of galaxies are both important probes to measure the growth of
cosmic structures of the Universe and excellent astrophysical
laboratories \citep[see][and references therein]{kravtsov12}.  Their
measured number density and evolution strongly depend on the
cosmological parameters, particularly at the high mass end
\citep[see][]{allen11}.  Their number density evolution on a time
scale of $\sim 9$ Gyr, corresponding to the highest redshift where
virialized clusters are currently found ($z\geq 1.3$), ranges between
one and more than three orders of magnitudes for masses $M > 5 \times
10^{14} h^{-1} M_\odot$, depending on the cosmological parameters
\citep[see][]{2002Rosati}.  Therefore, the density of massive clusters
at high redshift allows one to trace the growth rate of perturbations,
thus providing a robust constraint for cosmological models.  Other
cosmological tests with galaxy clusters are provided by the power
spectrum of galaxy clusters \citep{balaguera11} and the baryonic
fraction \citep{allen02,ettori03,ettori09}.  The latter is a
geometrical test that relies on the assumption of the constancy of
the baryonic fraction as a function of the cosmic epoch and therefore
does not depend on the density evolution of galaxy clusters.

There is now a general convergence towards the standard $\Lambda$ cold dark matter
model ($\Lambda$CDM), which is based on observables related to the cosmic 
microwave background \citep[CMB,][]{komatsu11}, the
correlation properties and the growth rate of the large-scale structure of
the Universe \citep{reid12}, and the supernova (SNae) surveys \citep[for
  recent developments see][]{suzuki12}.  Although CMB and SNae
experiments that directly probe the geometry of the Universe have gained
large visibility, significant efforts are still being devoted to
cosmological tests with galaxy clusters thanks to the high
complementarity with the other tests \citep{mantz10}.  Recently, the
emphasis has shifted towards the search of deviations from the reference
$\Lambda$CDM model, like tests of general relativity
\citep{schmidt09,rapetti12}, coupled dark energy models
\citep{amendola12}, or non-Gaussianity in the primordial density
perturbation field \citep[e.g.,][]{sartoris10,sawa12}.  Potentially,
the mere presence of few massive, high-z galaxy clusters such as those
recently discovered \citep[e.g.,][]{jee09,foley11,santos11} could
create some tension with the standard $\Lambda$CDM and quintessence
models
\citep[e.g.,][]{jimenez09,baldi11,chong12,mortonson11,hoyle11,harrison12,waizmann12}.
So far, no evidence for such a discrepancy has been found, except when
considering the combined probability of the most massive, high-z
clusters \citep{jee11}.

X-ray selected clusters play a key role in this context.  Thanks to
the X-ray thermal emission from the intracluster medium (ICM), which
is the largest baryonic component of galaxy clusters, it is possible
to measure the cluster mass with good accuracy up to high-z.
Unfortunately, X-ray studies of distant galaxy clusters are limited
not only by relatively small samples, but also by low fluxes and
surface brightness,
which hamper a meaningful X-ray spectral analysis in most 
cases.  This can be achieved only with very deep exposures, possibly
with the high imaging quality that only \textit{Chandra} can reach.
An independent strategy is to find clusters and estimate their
masses through the Sunyaev-Zeldovich (SZ) effect.  Surveys carried out with
the Atacama Cosmology Telescope \citep{sehgal11} and the South Pole
Telescope \citep{reichardt12} are delivering the first remarkable
results.  However, the calibration of mass proxies for SZ-selected
clusters still relies on a cross-calibration with X-ray data
\citep{andersson2012,ade2012}.  Before SZ mass measurements are
robustly calibrated, the two major existing X-ray facilities,
XMM-Newton and {\it Chandra}, still offer the best way to find and
characterize distant massive clusters.

Significant progress in this field can be pursued with a two-step
strategy.  The first step consists of exploiting the {\it Chandra} and
XMM-Newton archives to identify high-z cluster candidates.  The total
solid angle covered by both telescopes amounts to about 100-200
deg$^2$ (excluding the Galactic plane), and only a minor fraction of
these fields reaches the required sensitivity to detect clusters at
$z>1$, whose typical flux is about $10^{-14}$ erg s$^{-1}$ cm$^{-2}$.
This has been best achieved by a number of ongoing surveys, namely, the XMM Distant
Clusters Project \citep[XDCP,][]{2011fassbender}, by far the most
sucessful to date, the XMM-Cluster
Survey \citep[XCS,][]{2011lloyd-davies}, and the XMM Large Scale
Structure Survey \citep[XMM-LSS,][]{2011adami}\footnote{For a quick overview of
  current X-ray surveys, see \citet{2012tundo}}.  As a second step,
it is necessary to perform in-depth studies of the most massive,
distant cluster candidates.  This strategy requires time-expensive
observations with \textit{Chandra} in order to obtain good signal-to-noise ratio (S/N) spectra
(typically collecting a total of 1000 net photons or more in the
0.5-7 keV band) and high-resolution images to remove the effect of
contaminating active galactic nuclei (AGN) emission and of central cool cores.  Clearly, this
goal can also be pursued with deep X-ray follow-up of SZ-selected
targets.

If we focus on the most distant cluster candidates, we find that only
nine clusters have been spectroscopically confirmed at $z\ge 1.5$ to
date and only some of them have estimated masses in excess of
$10^{14}$ $M_\odot$
\citep{brodwin11,brodwin12,fassbender11,nastasi11,santos11,papovich10,gobat11,zeimann12,stanford2012}.
We note that most of them have been identified in infrared (IR) surveys or by combining
near infrared (NIR) and X-ray data\footnote{No SZ-selected clusters are currently
  found above $z=1.5$, the most distant confirmed SZ-selected cluster
  is reported at $z\sim 1.32$ \citep{stalder12}.}.  Moreover, the apparent lack of
extended X-ray emission in some of these systems suggests low X-ray
luminosities and points towards a re-classification of these objects
as protoclusters. Typically, data used for detection do not allow
one to determine the virialization status of a structure, or to
measure the average temperature of the ICM.

In this paper, we present the identification and characterization of
the most distant X-ray cluster discovered with \textit{Chandra} at
$z\sim 1.5$ in the same field of the cluster WARPJ1415.1+3612
\citep[hereafter WARPJ1415, at $z_{spec}=1.03$, see][]{santos12}.
Since these X-ray data are the deepest observation ever carried out
with \textit{Chandra} on clusters at $z>1$ (370 ksec), they allowed us
not only to firmly detect the extended ICM emission, but also to
measure its properties from the X-ray spectrum.  In addition to the
X-ray data, we have high-quality multi-band optical/NIR data from
Subaru and infrared \textit{Spitzer}-IRAC (3.6$\mu m$) imaging.
Currently, the only comparable systems in terms of redshift and
estimated mass are XMMUJ2235.3-2557 at $z=1.393$ \citep{2009Rosati}, XMMUJ0044 at
$z=1.578$ \citep{santos11}, and, recently, IDCS J1426+3508 at $z=1.75$ 
\citep{stanford2012}. Only the first one, however, was observed with
sufficiently deep {\it Chandra} and XMM-Newton data to derive a robust
total mass estimate.

The paper is organized as follows.  In Section 2 we describe our
X-ray spectral analysis, which provides us with X-ray redshift,
temperature, and metallicity.  We also measure the surface brightness
profile and concentration and derive the ICM mass and the total
cluster mass.  In Section 3 we describe the optical and IR data,
discuss the red sequence, and compute galaxy photometric redshifts.  In
Section 4 we discuss the properties of CXO J1415.2+3610 in a
cosmological context and in comparison with other distant X-ray
clusters.  Finally, our conclusions are summarized in Section 5.
Throughout the paper, we adopt the seven-year WMAP cosmology
($\Omega_{\Lambda} =0.73 $, $\Omega_m =0.27$, and $H_0 = 70.4 $ km
s$^{-1}$ Mpc$^{-1}$ \citep{komatsu11}.  In this cosmology, $1 \arcsec$
on the sky corresponds to about $8.56$ kpc at $z = 1.5$.  Quoted
errors always correspond to a 1 $\sigma$ confidence level.  Filter
magnitudes are presented in the AB system, unless stated otherwise.

\section{Cluster detection and X-ray data analysis}

\subsection{X-ray data}

\begin{figure}
\includegraphics[width=8.5cm,angle=0]{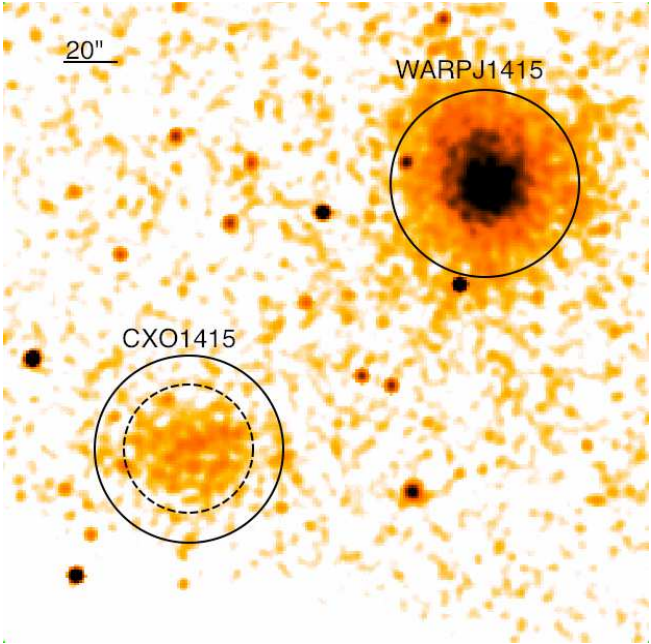}
\caption{ACIS-S+I image of the clusters field, smoothed with a $3
  \arcsec$ Gaussian kernel.  The bright cluster WARPJ1415 at $z=1$ is
  visible in the upper right, whereas CXO1415 lies at $\sim 2 \arcmin$
  in the south-west direction.  The solid circles have a radius of
  $35 \arcsec$ (corresponding to $300$ kpc at $z=1.46$), whereas the
  dashed circle with $r =24 \arcsec$ refers to the aperture of maximum
  S/N used for the spectral analysis.  The image has a size of $ 4
  \arcmin \times 4 \arcmin$.  }
\label{xray}
\end{figure}

CXO J1415.2+3610 (hereafter CXO1415, RA: 14:15:20.143,
DEC:+36:10:27.56) was serendipitously detected in a deep
\textit{Chandra} GO observation of 280 ks with ACIS-S granted in
Cycle 12 (PI J Santos) to study the ICM properties of the cool-core
cluster WARPJ1415 at $z=1.03$.  The new system, located at a distance
of $\sim 2$ arcmin in the south-west direction of WARPJ1415, was
immediately detected by visual inspection as an extended X-ray source
in the ACIS-S image (see Figure \ref{xray}).  The cluster was also
visible in a shallower ACIS-I image of 90 ks available in the
\textit{Chandra} archive, which we also use in this study.  We 
tried to recover the XMM data for WARPJ1415 to extract more
information on CXO1415.  Unfortunately, CXO1415 is barely visible in
the total XMM image (PN plus 2 MOS) and is partially affected by a
gap.  The XMM exposure time after cleaning from high-background
interval is about 16 ks.  Therefore, XMM data do not provide any
useful signal, and our analysis is entirely based on the
\textit{Chandra} data.

The total exposure time of the analyzed data is 370 ks
(ACSI-S+ACIS-I).  The data reduction and source extraction were
performed with CIAO 4.3 and CALDB v4.4.5.  A detailed description of
the data reduction procedures can be found in \citet{santos12}.  The
circular region of maximum S/N is obtained for
a radius of 24 arcsec centered on RA=14:15:20.143, DEC=+36:10:27.56.
The source is detected with a S/N of $11.6$ and $\sim 3$, corresponding
to about 540 and 115 net counts (0.5-7 keV band) in ACIS-S and
ACIS-I, respectively.

\subsection{Spectral analysis: temperature, metallicity, and redshift}

We perform a combined spectral analysis of the ACIS-S and ACIS-I
data with {\tt Xspec} v12.6 \cite{arnaud96}.  The adopted spectral
model is a single-temperature {\tt mekal} model \citep{kaa92,lie95},
using as a reference the solar abundance of \citet{asp05}.  The local
absorption is fixed to the Galactic neutral hydrogen column density
measured at the cluster position and equal to $N_{H}= 1.05 \times 10
^{20}$cm$^{-2}$ taken from the LAB Survey of Galactic HI
\citep{2005LAB}.  The fits are performed over the energy range
$0.5-8.0$ keV.  We used Cash statistics applied to the source plus
background, which is preferable for low S/N spectra \citep{nou89}.

\begin{figure}
\includegraphics[width=8.5cm,angle=0]{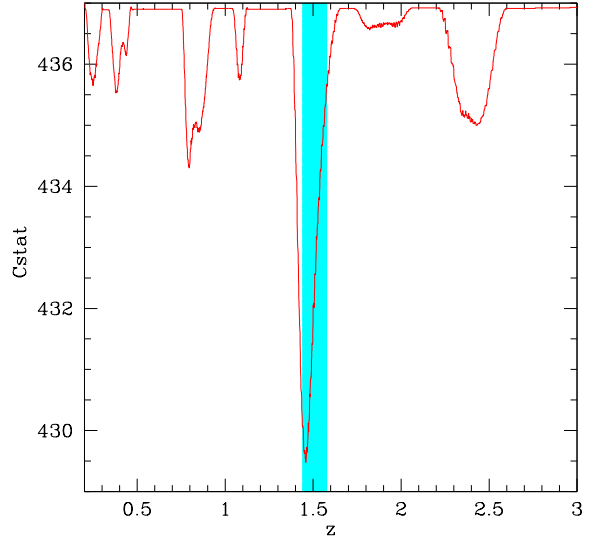}
\caption{$C_{stat}$ as a function of redshift parameter for the
  combined (ACIS-S+ACIS-I) spectral analysis of CXO1415.  The
  minimum is obtained for $z=1.46$, when $\Delta C_{stat} \sim 8$.
  The cyan stripe shows the 1 $\sigma$ confidence interval obtained
  from the photo-$z$ ($z_{phot} = 1.52 \pm 0.06$, see Section 3.2)}
\label{cstat}
\end{figure}

First, we search for the $K_\alpha$ Fe line complex in order to
measure the redshift $z_X$ directly from the X-ray analysis.  For
\textit{Chandra} observations, about 1000 net counts are needed to
measure $z_X$ at a $3\sigma$ confidence level, and even more are needed in the
case of hot clusters, as shown in \citet{2011Yu}.  Nevertheless, we
run a combined fit on the ACIS-S and ACIS-I spectra, leaving free
redshift, temperature, metal abundance, and normalization.  We find
that the best fit gives $z_X = 1.46 \pm 0.025$.  To better quantify the
robustness of the detection of the Fe $K_\alpha$ line complex, we plot in
Figure \ref{cstat} the quantity $C_{stat}$ as a function of
the redshift parameter, while all the other spectral parameters are
set to the best-fit values.  The best-fit redshift is obtained with a
$\Delta C_{stat}\sim 8$.  Formally, a 3 $\sigma$ confidence level is
obtained when $\Delta C_{stat}= 9$ \citep[see][]{2011Yu}.  As a result,
the redshift measured from the X-ray analysis corresponds to a $\sim
2.8 \sigma$ (99.5\% confidence level) detection of the Fe line.
However, as we show later, another relevant piece of information
comes from the photometric redshift, which allows us to exclude
secondary minima at $z\sim 0.8$ and $z\sim 2.5$ and increases the
robustness of the redshift value obtained from the X-ray analysis.
Therefore, in the following analysis we assume $z=z_X=1.46$.  
This result shows that the measure of the redshift
from X-ray spectral analysis is possible up to the highest redshift
where X-ray clusters are currently detected.  This aspect is extremely
important when designing a wide and deep survey mission for
the future, such as the Wide Field X-ray Telescope
\citep{2010WFXT,2010WFXT2,2010WFXTsimu}.

\begin{figure}
\hspace{0.5cm}
\includegraphics[width=6.0cm,angle=-90]{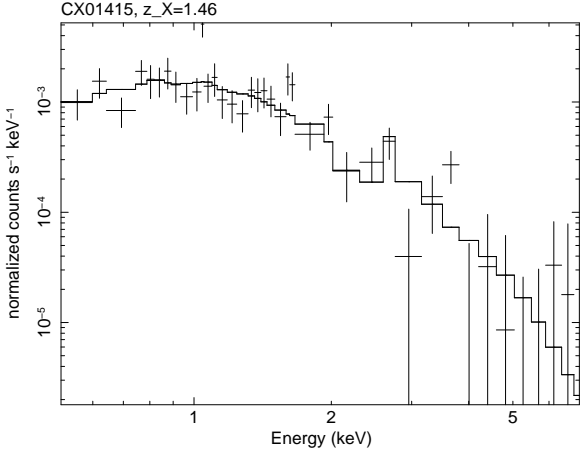}
\caption{Folded \textit{Chandra} ACIS-S spectrum of CXO1415 with its
  best-fit {\tt mekal} model (solid line). The iron $K_\alpha$ line
  complex is detected at the 99.5\% confidence level at 2.8 keV in
  the observed reference frame.}
\label{spectrumS}
\end{figure}

The cluster unabsorbed soft-band flux within a circular region of
$24\arcsec$ radius is $S_{0.5-2.0 keV}=(6.4 \pm 0.8) \times 10^{-15}$
erg s$^{-1}$ cm$^{-2}$, and the hard flux $S_{2-10 keV}=(5.8 \pm 0.7)
\times 10^{-15}$.  The cluster emission-weighted temperature is $kT=
5.8_{-1.0}^{+1.2}$ keV.  The measured Fe abundance, in units of
\citet{asp05}, is $Z_{Fe} = 1.3_{-0.5}^{+0.7} Z_{Fe\odot}$.  We warn
that Fe abundance values measured freezing the redshift parameter to
the $z_X$ value are slightly biased high \citep[see][]{2011Yu}.  Finally, we
note that this analysis is in very good agreement with that based on
the ACIS-S data only, which indeed carries most of the signal.  The
inclusion of the ACIS-I data helps to slightly reduce the statistical
error bars.  For sake of clarity, we
show in Figure \ref{spectrumS} the spectrum of CXO1415 obtained only with ACIS-S, along with
the best-fit model.

\subsection{Surface brightness profile and concentration}

\begin{figure*}
\begin{center}
\hspace{-2.cm}
\includegraphics[width=8.0cm,angle=0]{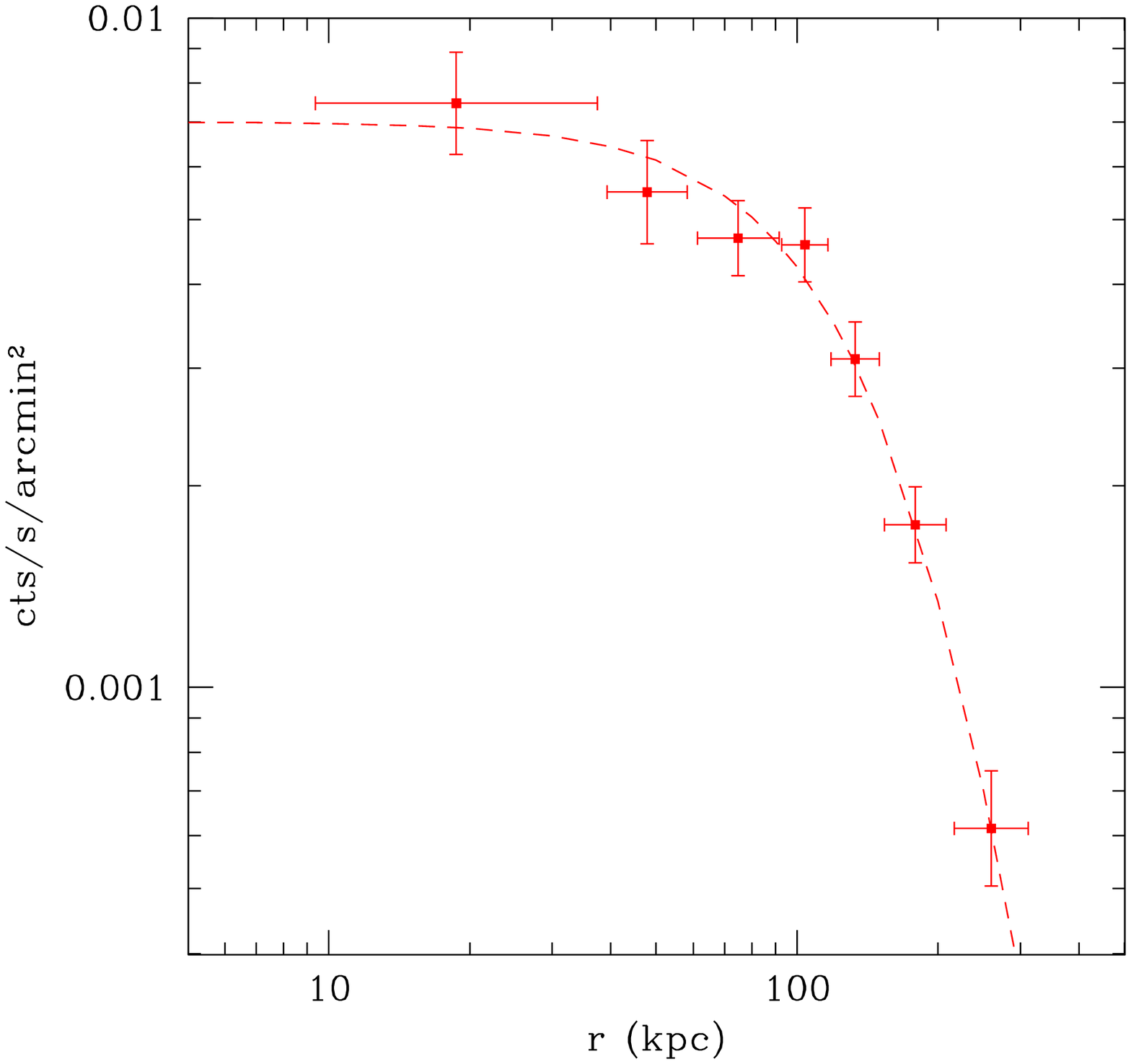}
\hspace{1.cm}
\includegraphics[width=8.0cm,angle=0]{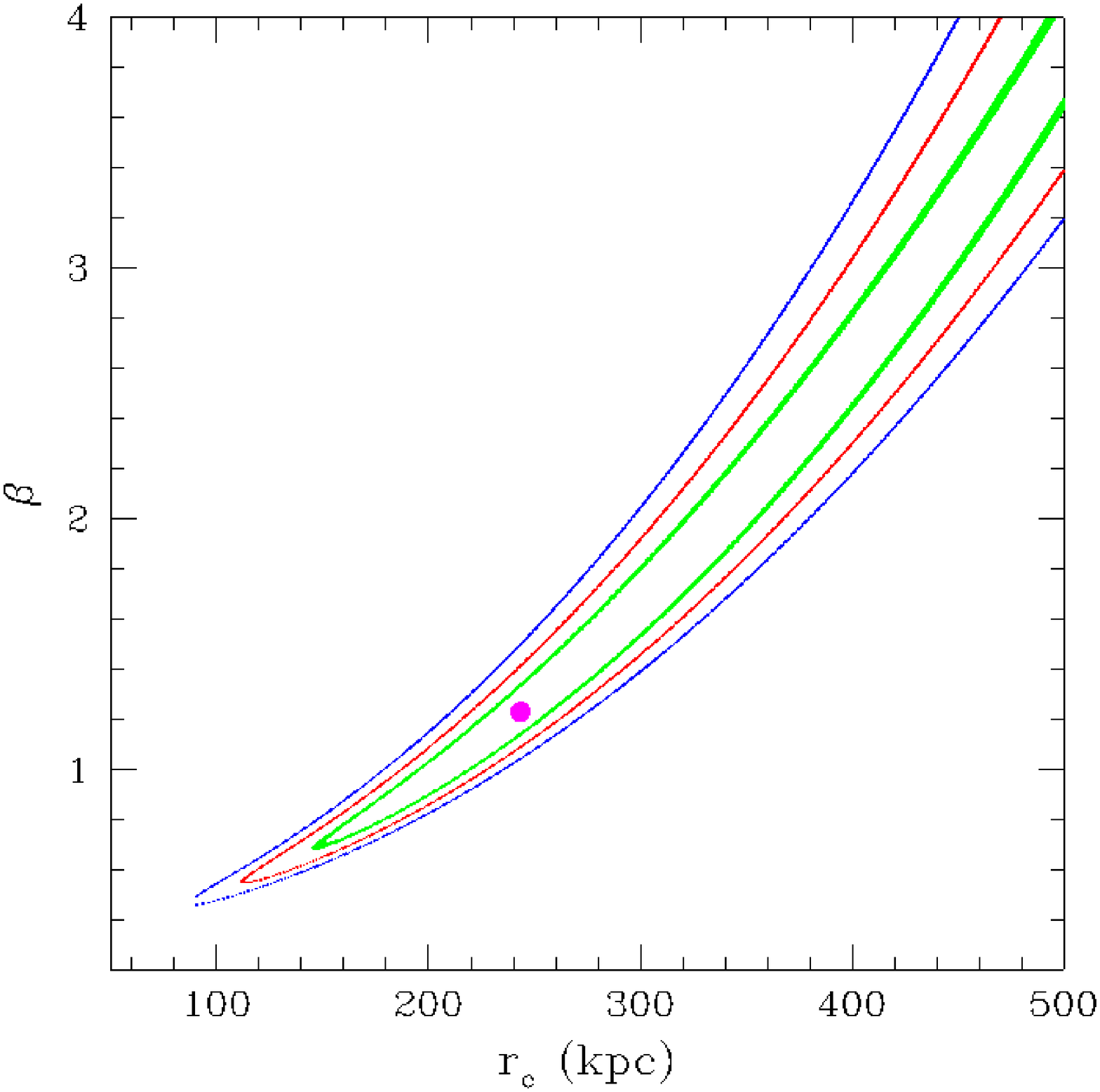}
\hspace{1.5cm}
\end{center}
\vspace*{-0.4cm}
\caption{{\it Left}: Background-subtracted surface brightness profile
  in the soft (0.5-2 keV) band (points) and best-fit $\beta$ model
  (red dashed line) for CXO1415.  Error bars correspond to 1 $\sigma$
  uncertainty.  {\it Right}: 1-2-3 $\sigma$ confidence levels for two
  relevant parameters in the $\beta$-$r_c$ parameter space.}
\label{SB}
\end{figure*}

To measure the surface brightness of CX01415 we use only the ACIS-S
image.  The azymuthally averaged surface brightness profile, which was computed
in the exposure-corrected ACIS-S image in the soft band, is traced
out to about $35\arcsec$ and is well described by a single $\beta$
model, $S(r) = S_{0} (1+(r/r_{c})^2) ^{-3\beta+0.5} + bkg$
\citep{cav76}.  The best-fit model, shown in Figure \ref{SB}, left-hand
panel, has a slope $\beta=1.2$ and a core radius $r_{c} = 240 $ kpc
for a reduced $\tilde \chi^2 = 0.7$.  The uncertainties on the profile
parameters are quite large, given the degeneracy between $\beta$ and
$r_c$ (see Figure \ref{SB}, right-hand panel).  This degeneracy does not
significantly affect the overall profile within the extraction radius,
which is used to derive both total and baryonic mass within 300 kpc in
Section 2.4.  However, the extrapolation of the surface brightness
profile at larger radii is uncertain, and further assumptions are
needed to estimate the total mass up to the virial radius.  The
total luminosity, being dominated by the inner regions, will be mildly
affected.

The luminosity emitted within 24 arcsec in the soft band is
$L_{X}[0.5-2.0]=(0.61 \pm 0.07) \times 10^{44}$ erg s$^{-1}$ and the
bolometric luminosity $L_{bol} =(2.1 \pm 0.2) \times 10 ^{44}$ erg
s$^{-1}$.  To compute the total cluster rest-frame luminosity, we
apply a correction factor obtained by extrapolating the fit of the
surface brightness profile up to the virial radius (see Section 2.4).
We obtain a total soft-band luminosity of $L_{X}[0.5-2.0]=(0.93 \pm
0.10) \times 10^{44}$ erg s$^{-1}$ and a bolometric $L_{bol} =(3.2 \pm
0.3) \times 10 ^{44}$ erg s$^{-1}$.  This is probably an upper limit,
given that surface brightness profiles are generally observed to steepen
in the outer regions with respect to the extrapolation of
the beta-fit model \citep{ettori09b}.

The rather flat surface brightness profile within 100 kpc does not
support the presence of a cool core.  To quantify the core strength,
we compute the value of $c_{SB}$, a concentration parameter defined in
\citet{santos08} as the ratio of the surface brightness within $40$
kpc and $400$ kpc: $c_{SB}=SB\,(<40kpc)\,/\,SB\,(<400kpc)$.  A cluster
is expected to host a cool core if $c_{SB} > 0.075$.  This simple
phenomenological parameter has proven to be a robust cool core
estimator, particularly in low S/N data typical of distant clusters.
We measured $c_{SB}=0.059 \pm 0.017$, thus confirming that CXO1415
does not show evidence of a well developed cool core.

\subsection{Mass measurement}

The baryonic mass can be directly computed once the three-dimensional
electron density profile is known.  For a simple beta model, this is
given by $n_e(r) = n_{e0}[ (1+(r/r_{c})^2) ^{-3\beta/2}]$.  To measure
$n_{e0}$, we use the relation between the normalization of the X-ray
spectrum and the electron and proton density in the ICM for the {\tt
  mekal} model:

\begin{equation} 
Norm = {{10^{−14}}\over { 4\pi \, D^2_a\, (1 + z)^2}}\times \int
n_en_HdV \, ,
\end{equation}

\noindent
where $D_a$ is the angular diameter distance to the source (cm), and
$n_e$ and $n_H$ (cm$^{-3}$) are the electron and hydrogen densities,
respectively, and the volume integral is performed over the projected
region used for the spectral fit ($24$ arcsec).  For CXO1415, this
gives a central density of $n_{e0} = (0.99 \pm 0.13) \times 10^{-2}
$cm$^{-3}$.

The total cluster mass can be estimated under the assumptions of
hydrostatic equilibrium and spherical symmetry, which leads to the
simple equation \citep{sarazin88}

\begin{equation}
M(r) = -4.0\times 10^{13} M_\odot T (keV) r (Mpc) \Bigg( {{dlog(n_e)}\over{
    dlog r}} + {{dlog(T)}\over{dlog r}}\Bigg) \, .
\end{equation}

Since we are not able to measure the temperature profile, we assume
isothermality inside the extraction radius and beyond that adopt a
mildly decreasing temperature profile $kT\propto r^{-0.24}$, as found
in local clusters \citep[see][]{leccardi08}.  Therefore the term in
parenthesis in Equation (2) is contributed by a term due to the slowly
declining temperature profile and by the slope of the density profile,
which is simply $ -3 \, \beta \, x^{2} / (1+x^{2})$, where $x =
r/r_{c}$.  Since we can robustly measure the slope of the density
profile only up to 300 kpc, we decide to keep the slope constant
beyond 300 kpc (${{dlog(n_e)}/{ dlog r}} = 2.2$).

To measure the total mass at a given density contrast, we solve the
equation $M_\Delta(r_\Delta) = 4/3\pi r_\Delta^3 \, \Delta
\rho_c(z_{cl})$ to compute the radius where the average density level
with respect to the critical density $\rho_c(z_{cl}) $ is $\Delta$.
Typically, mass measurements are reported for $\Delta = 2500, 500$, and
$200$.  The 1$\sigma$ confidence intervals on the mass are computed by
including the error on the temperature and on the gas density profile.
The results are shown in Table \ref{mtab}.  Only the radius
$r_{2500}=(220\pm 55)$ kpc is well within the detection region.  We
obtain $M_{2500} = 0.86_{-0.17}^{+0.21} \times 10^{14} M_\odot$.  We
also report the total mass within 300 kpc $M (r< 300 kpc) =
1.38_{-0.28}^{+0.33} \times 10 ^{14} M_\odot$ and the corresponding
ICM mass $M_{ICM}(r<300 kpc) = 1.09_{-0.20}^{+0.30}\times 10^{13}
M_\odot$.

The total mass extrapolated to $r_{500}$, with the additional
assumption about the temperature and density profiles, are $M_{500} =
2.07_{-0.50}^{+0.70} \times 10 ^{14}M_\odot$ for $R_{500} =
510_{-50}^{+55}$ kpc.  We also extrapolate the mass measurement up to
the nominal virial radius, finding $M_{200} = 2.8_{-0.7}^{+1.0} \times
10 ^{14}M_\odot$ for $R_{200} = 760_{-70}^{+80}$ kpc.  The ICM mass
fraction is almost constant with a value $f_{ICM} = 0.10 \pm 0.02$,
which is in very good agreement with the empirical relation between
$f_{ICM}$ and $M_{500}$ shown in \citet{2009Vikhlinin}.

Finally, for completeness, we recompute the masses after relaxing the
assumption of isothermality within the extraction radius and adopting
the temperature profile $kT\propto r^{-0.24}$ at any radius.  In this
case, the only requirement is that the average temperature within this
radius is $5.8$ keV as observed.  With these assumptions, masses at
$M_{500}$ and $M_{200}$ are $\sim 18$\% lower than those obtained with
the simplest choice $T=const$ for $r<r_{ext}$.  This difference should
be regarded as systematic, since the temperature profile within the
extraction radius may well behave differently from a simple power law,
with significantly different effects on the extrapolated masses.  We
prefer not to consider this value in the following discussion.

\begin{table}
 \centering
\caption{Summary of the mass estimates for CXO1415.  The ``*''
  indicates that the mass values are extrapolated beyond 300 kpc, the
  maximum radius where the ICM emission is actually detected.}
\label{mtab}
\begin{tabular}{@{}cccc@{}}\hline
$\Delta$ & Radius & $M_{tot}$  &     $M_{ICM}$     \\ 
         &  kpc   & $M_\odot$  &     $M_\odot$     \\ 
\hline  \\ 
$2500$  & $R_{2500}=220\pm 55$ & $0.86_{-0.17}^{+0.21} \times 10^{14}$ & $0.64_{-0.13}^{+0.16} \times 10^{13}$  \\ 
 \\ 
  -  & $R_{ext}=300$ & $1.38_{-0.28}^{+0.33} \times 10 ^{14}$ & $1.09_{-0.20}^{+0.30}\times 10^{13}$    \\ 
 \\ 
  $500^*$  & $R_{500} = 510_{-50}^{+55}$ &  $2.07_{-0.50}^{+0.70} \times 10 ^{14}$  & $2.13_{-0.50}^{+0.70} \times 10^{13}$  \\ 
 \\ 
$200^*$  & $R_{200} =  760_{-70}^{+80}$ & $2.8_{-0.7}^{+1.0} \times 10 ^{14}$ & $2.9_{-0.7}^{+1.0} \times 10^{13}$     \\ 
 \\ 
\hline
\end{tabular}
\end{table}

\subsection{Systematics in the total mass measurements}

\begin{table}
 \centering
\caption{Total mass at $R_{500}$ and $R_{200}$ obtained under
  different assumptions for CXO1415. (2) \citet{2009Vikhlinin}; (3a)
  \citet{2009Vikhlinin}; (3b) \citet{fabjan11}; (4) \citet{nfw96}.}
\label{mtab_extrap}
\begin{tabular}{@{}cccc@{}}\hline
Method & $R_{500}$    & $M_{500}$   & $M_{200}$   \\ 
             & kpc             & $M_\odot$   & $M_\odot$   \\ 
\hline  \\ 
M-T relation (2) & - & ($2.4\pm 0.6)\times 10^{14}$ & -  \\ 
 \\ 
$Y_X-M$ relation  (3a) & $495 \pm 45$ & $1.93_{-0.43}^{+0.55}\times 10^{14} $  \\
 \\ 
$Y_X-M$ relation  (3b) & $560_{-40}^{+50}$ & $2.8_{-0.6}^{+0.8}\times 10^{14}$  & -  \\ 
 \\ 
NFW (4)  & $540 \pm 90$ & $(2.6\pm 1.8)\times 10^{14}$ & $(3.7\pm 2.0)\times 10^{14}$ \\ 
  \\ 
\hline
\end{tabular}
\end{table}

The mass measurements at $R_{500}$ and $R_{200}$ have been
obtained under the assumption of hydrostatic equilibrium as an
extrapolation of the observed profile beyond the largest radius where
X-ray emission is detected ($300$ kpc), as well as under some reasonable
assumptions on the slope of the ICM density and temperature profiles
beyond this radius.  Our assumptions are: i) the slope of the
density profile remains constant, $\sim 2.2$ beyond 300 kpc; ii) the
temperature profile declines as $\propto r^{-0.24}$ outside the
extraction radius as observed in local clusters in the $(0.1-0.6)
\times r_{180}$ range \citep[][]{leccardi08}.  In order to evaluate
possible systematics associated to the extrapolation of the total mass
to $R_{500}$ and $R_{200}$, we present the mass estimates adopting
three different methods.

The first estimate (method 2) is based on the redshift-dependent
scaling relations calibrated on local clusters and presented in
\citet{2009Vikhlinin}.  From the empirical relation described in their
Table 3, we find $M_{500} = (2.4 \pm 0.6) \times 10^{14} M_\odot$.  As
a third method, we use the integrated Compton parameter $Y_X \equiv T_X
\times M_{ICM}$, which is considered a robust mass proxy within
$R_{500}$, as shown by numerical simulations
\citep[e.g.,][]{kravtsov06}.  The observed value for CXO1415, assuming
$r_{500}$ as obtained from method 1, is $Y_X =
1.24^{+0.48}_{-0.36}\times 10^{14}$ keV $M_\odot$.

If we use the $Y_X$-$M$ empirical relation taken from Table 3 of
\citet{2009Vikhlinin}, we obtain $R_{500} = (495 \pm 45)$ kpc and
$M_{500} = 1.93_{-0.43}^{+0.55} \times 10^{14} M_\odot$.  Using the
calibration based on numerical simulations obtained in
\citet{fabjan11}, we find $M_{500} = 2.8_{-0.6}^{+0.8} \times
10^{14} M_\odot$, for $R_{500} = 560_{-40}^{+50} $ kpc.  Both values
are obtained iteratively in order to compute consistently all the
quantities at $r_{500}$.  The discrepancy of a factor $\sim 1.5$
between the numerical simulations and the phenomenological
$M_{500}$-$Y_X$ relation can be partially explained with some 
nonthermal pressure support not accounted for in the hydrostatic mass
estimates \citep[see][]{kravtsov06}.

We also extrapolate the mass according to the Navarro, Frenk \& White 
profile \citep[NFW,][method 4]{nfw96}, after requiring a normalization at 300 kpc
to the total mass value actually measured from the data.  Still, the
extrapolation depends on the unknown concentration parameter.
However, such dependence is not strong and can be included in the
extrapolated value once we restrain the concentration in the plausible
range $c_{NFW} = 4.0 \pm 0.5$, as found in simulations for the wide
mass range $5\times 10^{13} M_\odot< M < 5\times 10^{14}M_\odot$
\citep{gao08}.  We find $M_{500} = 2.6^{+1.3}_{-0.9} \times
10^{14}M_\odot$ and $M_{200} = 3.7^{+2.0}_{-1.2} \times
10^{14}M_\odot$.  Here, the much larger error bars include both the
statistical error on the measured mass at $R=300$ kpc and the
uncertainty on the concentration parameter.

The systematic uncertainty due to the different mass measurement
methods can be appreciated in Table \ref{mtab_extrap}.  We find that
the systematics error on $M_{500}$ is about 20\%, slightly lower than
the $1 \sigma$ confidence level associated to the statistical
uncertainty, which is $\sim 30$\%.  We note that mass estimates based
on a beta-profile fitting and the assumption of the hydrostatic
equilibrium can give values $\sim 20$\% lower around $r_{500}$ (and
even larger for larger radii) due to the violation of hydrostatic
equilibrim and the presence of significant bulk motions in the ICM.
This has been shown in numerical simulations
\citep[see][]{bartelman96,rasia04,borgani04}.  Observational
calibrations also show that hydrostatic masses underestimate weak
lensing masses by 10\% on average ar $r_{500}$
\citep[see][]{hoekstra07,mahdavi12}.  We do not find evident bias when
comparing our hydrostatic mass (method 1) with values obtained with
empirical calibrations (methods 2, 3), also because the statistical
errors are larger than the possible effects of residual bulk motions.
Therefore, all the mass measurements we obtained are consistent with
each other within the errors.  Our conclusion is that it is feasible to measure
$M_{500}$ for high-z, massive clusters using deep {\it Chandra}
exposures, although clusters with fluxes below $10^{-14}$
erg s$^{-1}$ cm$^{-2}$, like CXO1415, suffer a nonnegligible
systematic uncertainty due to some unavoidable extrapolation.

\section{Optical/IR analysis}

The field of CXO1415 has wide optical photometric coverage with
Subaru-\textit{Suprime} (BVRiz), in addition to near-IR JK$_s$ with
Subaru-\textit{Moircs} and Spitzer-IRAC 3.6$\mu$m.  A
summary of the archival photometric data used in our analysis is
provided in Table \ref{opttab}.  Standard reduction procedures using
IRAF routines were used to produce the optical/NIR science images
retrieved from SMOKA \citep{baba02}, with an average seeing of
$0.68\arcsec$ in the optical bands, and $0.60\arcsec$ in the near-IR.
Finally, we used level 2 Spitzer-IRAC 3.6 $\mu$m
observations from the archive, which are higher level
reduced (pbcd) products.

\begin{figure*}
\begin{center}
\includegraphics[width=9.5cm,angle=0]{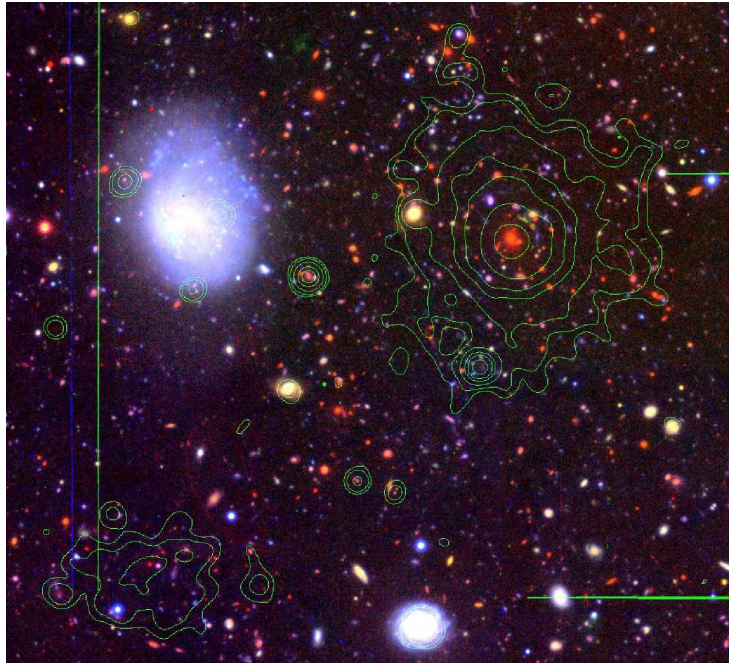}
\end{center}
\vspace*{-0.4cm}
\caption{Field color image (Suprime-Cam BRZ filters) of WARPJ1415 (top right) and
  CXO1415 (bottom left) with soft X-ray contours overlaid in green.  The image is 3.5 arcmin 
  horizontally across.}
\label{color}
\end{figure*}

\begin{figure*}
\begin{center}
\hspace{-2.cm}
\includegraphics[width=9cm,angle=0]{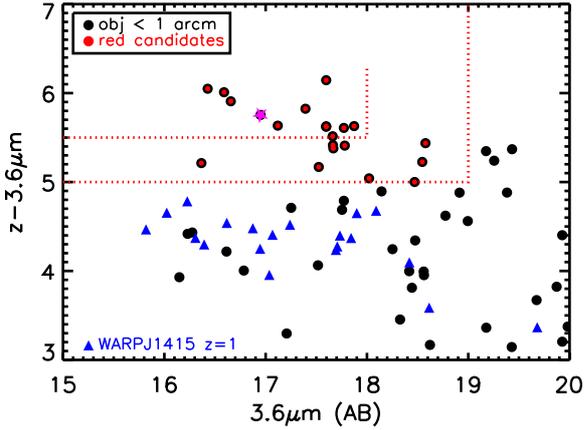}
\hspace{1.cm}
\includegraphics[width=6.5cm,angle=0]{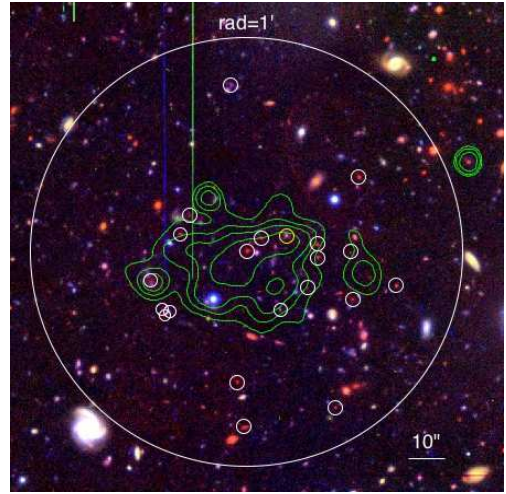}
\hspace{1.5cm}
\end{center}
\vspace*{-0.4cm}
\caption{\textit{Left}: ($z-3.6 \mu$m) color-magnitude relation of
  CXO1415. Only objects within $1 \arcmin$ distance from the X-ray
  center are shown (black circles). The 21 red cluster candidates (red
  circles) are selected on the basis of the color cut $5< z-3.6 \mu m
  <6.5$.  For comparison we also plot 19 confirmed members of
  WARPJ1415 at $z=1$ (blue triangles).  The magenta star indicates the
  galaxy coincident with the X-ray peak.  \textit{Right} Selection of
  red, high-z cluster galaxy candidates (white circles), with X-ray
  contours displayed in green. The brightest candidate is marked by a
  yellow circle.}
\label{cmd}
\end{figure*}

\begin{table}
 \centering
\caption{Summary of the optical/NIR/IR imaging datasets. BVRiz are
  from Subaru-\textit{Suprime}, JK$_s$ from Subaru-\textit{Moircs}, and
  3.6$\mu$m from Spitzer-IRAC.}
\label{opttab}
\begin{tabular}{@{}cccc@{}}\hline
\\
Band        & Exposure       & Date obs.   & Image quality \\ 
            &     (s)        &             &   (\arcsec)   \\ 
\hline 
\\
B           &  6$\times$360  & 2006-05-30  &  1.10         \\ 
V           &  6$\times$360  &  2003-04-05 &  0.66         \\ 
R           &  6$\times$480  &  2003-04-04 &  0.74         \\
i           &  3$\times$300  & 2001-06-27  &  0.70         \\ 
z           &  9$\times$180  & 2001-06-27  &  0.62         \\ 
J           &  2160	     & 2009-04-19  &  0.62         \\ 
K$_s$           &  1440          & 2009-04-19  &  0.56         \\ 
3.6$\mu$m   &  97            & 2006-02-05  &  1.7          \\
\\
\hline
\end{tabular}
\end{table}

\subsection{Color-magnitude diagram}

Relaxed galaxy clusters are characterized by a well-defined sequence
of red, passively evolving galaxies \citep[see][]{gladders05}.  
This relation has been shown to remain fairly constant up to $z=1.4$
\citep{mei09,strazzullo10,2009Rosati}.  However, recent studies at $z>1.4$
indicate the appearance of star formation in the central brightest 
galaxies \citep{hilton10,fassbender11,santos11},
with an increasing scatter in the red sequence.

In this section, we investigate the ($z'-3.6 \mu$m) color-magnitude
relation (CMR) of CXO1415.  This optical/IR diagnostic has proven to
be an accurate tool to study the galaxy population of $z>1$ clusters
\citep[e.g.,][]{muzzin09,brodwin10}.  The BRz color image reveals a
population of red galaxies centered on the X-ray contours, as shown in
Figure~\ref{color}.  We use SExtractor \citep{bertin96} to perform
source detection and galaxy photometry.  We measure aperture
magnitudes corrected to $1\arcsec$ in the z-band and apertures of
$1.8\arcsec$ in the $3.6\mu$m band. The total $3.6\mu$m magnitude is
given by the SExtractor parameter {\tt MAGAUTO}.  An aperture
correction corresponding to $2.5\times {\tt log}(1.112)$ was applied
to the IRAC band to match the optical
magnitudes\footnote{http://irsa.ipac.caltech.edu/data/SPITZER/docs/irac/iracinstrument
  handbook/28/\_Toc296497400}.

The $z'-3.6 \mu$m CMR, which includes only objects within a cluster-centric
distance of $1\arcmin$, is presented in Figure \ref{cmd}, left-hand panel.
The galaxy coincident with the X-ray peak (indicated by a magenta
star) has a total $3.6\mu$m magnitude of $16.9$ and a color equal to
$5.7$.  Contrary to what is commonly observed at lower redshifts, this
galaxy is not the brightest cluster galaxy.
A sequence of red galaxies clearly stands out at $5< z' -3.6 \mu$m
$<6.5$. For comparison we also plot the red sequence of 21 out of 25
spectroscopically confirmed members of WARPJ1415 \citep{huang09},
which has a median color of $4.4$ mag.  If we further restrict the
color cut to $5.5<z'-3.6\mu$m $<6.3$ and the total magnitude to below
$18$ mag, we obtain a more stringent locus for the red sequence,
described by a linear fit with a slope of $-0.25$. The mean
($z'-3.6\mu$m) color computed for the enclosed 12 galaxies is then
$5.8$ mag, with a spread of $0.21$ mag.  This already shows that
CXO1415 is at considerably higher redshift than 1.  The cluster
red sequence is well approximated by a \citet{bruzual03} single
starburst model with passively evolving galaxies with an age of 2
Gyr.  These models, which accurately reproduce the red-sequence of
WARPJ1415 at $z=1$, predict a redshift of $\sim 1.5$ for CXO1415 for a
color ($z'-3.6\mu$m$)=5.8$. This simple redshift estimate is
improved in the next section by using all eight photometric bands.

\subsection{Photometric redshifts}

\begin{figure}
\includegraphics[width=8.5cm,angle=0]{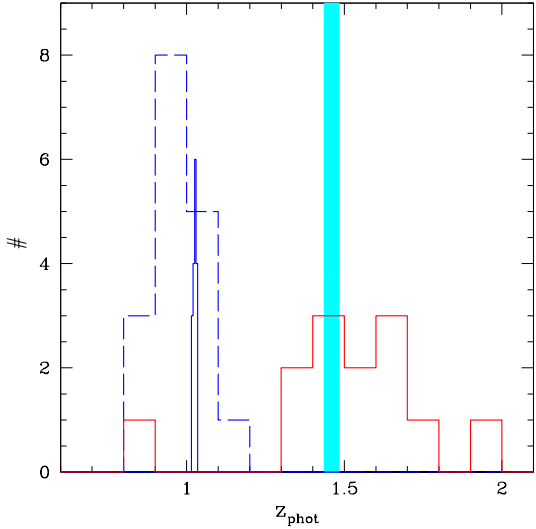}
\caption{Histogram of photometric redshifts computed in the cluster
  field. We show the photometric redshift distribution of 17
  spectroscopic members of WARPJ1415 at $z=1.03$ (blue dashed line)
  and the photometric redshift distribution of 13 candidate members
  selected from the red sequence of CXO1415 (red solid line).  The
  thinner histogram (solid blue line) marks the spectroscopic redshift
  distribution of WARPJ1415, while the cyan stripe corresponds to the X-ray
  redshift $z_X = 1.46\pm 0.025$.}
\label{photoz}
\end{figure}

Using our eight-band multicolor catalog, we computed photometric
redshifts for the candidate red galaxies with the photo-$z$ code
\textit{zebra} \citep{feldmann06} by comparing the observed magnitudes
with those expected from template spectral energy distributions (SED)
by \citet{bruzual03} through a standard SED fitting procedure.  Our
results for CXO1415 were calibrated against WARPJ1415 at $z=1.03$, for
which we obtained an accurate mean photo-$z$ of $\langle
z_{phot}\rangle =0.98$, with an rms dispersion $\sim 0.08$ based on
the colors of 17 spectroscopic members.  The comparison of
spectroscopic and photometric redshift for WARPJ1415 members shows
that the photo-$z$ accuracy on the average redshift is $\Delta
\langle z \rangle =0.05$.  Of the 21 CXO1415 galaxy candidates
selected in the CMR, four are outside the FoV of the MOIRCS data. A
visual inspection of these galaxies forced us to discard another four
galaxies because of their faint magnitudes and blending with nearby
objects, which may result in unreliable photometry.  The histogram of
the photometric redshifts is shown in Figure \ref{photoz}.  Based on
the SED fit of 13 red candidates, we find for CXO1415 $\langle
z_{phot}\rangle =1.52 \pm 0.06$, where the 1 $\sigma$ error is
estimated by extrapolating to $z=1.5$ the photo-$z$ accuracy measured
at $z=1$, assuming a dependence $\propto (1+z)$.  This result is in
good agreement with the $z_X$ obtained from the X-ray spectral
analysis.

At present, we do not yet have the spectral confirmation of the
redshift of CXO1415.  We recently analyzed long-slit spectra of two
galaxies (including the central one) obtained with OSIRIS
\citep[Optical System for Imaging and low-intermediate Resolution
  Integrated Spectroscopy,][]{cepa00} on the 10m facility Grantecan at
La Palma (Spain).  Unfortunately, we collected only one out of seven hours
originally requested, and the quality of the spectra was not
sufficient to derive reliable redshifts.  More efforts to achieve a
spectroscopic measure of the redshift of CXO1415 are ongoing.

Finally, we note that an in-depth morphological study of the cluster
galaxies would be extremely interesting.  Robust morphological
classification of cluster galaxy candidates would require HST/WFC3
imaging. Our Subaru NIR imaging in very good seeing data suggests that
about half of the selected red sequence galaxies show a distorted
late-type morphology, which is not uncommon in clusters at these
redshifts \citep[see][]{hilton10,fassbender11,santos11}.

\section{Discussion}

CXO1415 was originally discovered by visual inspection in the original
80 ks ACIS-I exposure.\footnote{This also suggests that a ``crowd-sourcing'' project of the kind of Galaxy Zoo
  \citep[see][]{lintott11} may be very useful if applied to {\it
    Chandra} X-ray images.}  In order to estimate  the
effective sky area within which it was found, we carried out a
simple experiment based on an inspection of the same image at
different exposure times.  This experiment allowed us to determine the
minimum exposure time necessary for a mere detection of the extended
emission.  By trimming the exposure time of the ACIS-S and ACIS-I
image of CXO1415, we found that 20 net counts can be considered a firm
limit for detection as an extended source via visual inspection.  This
implies that CXO1415 could have been detected in all the ACIS-I
observations with more than 15 ks and all the ACIS-S observations
with more than 12 ks.  The most conservative choice is to compute the
solid angle covered by all the Chandra ACIS-I and ACIS-S
extragalactic fields where CXO1415 may have been detected.  We browsed
the entire {\it Chandra} archive up to Cycle 12, combined multiple
exposures of the same region of the sky, and found that a total of
about $70$ deg$^2$ had been observed at the required depth.  This must
be considered as an upper limit, since we do not take into account the
solid angle lost due to the presence of bright extended targets in
several extragalactic {\it Chandra} fields.

We can use this extremely conservative choice of sky coverage to
estimate the maximum search volume from where CXO1415 has been drawn.
Using this information in combination with the cluster mass function
of \citet{tinker08} with WMAP7 parameters
\citep{komatsu11}\footnote{Specifically, we assumed a flat $\Lambda$CDM
  model with $\sigma_8=0.81$, $\Omega_0 = 0.275$, $h = 0.70$ for a
  flat geometry.}, we find that CXO1415 by itself, after considering
the uncertainty in its virial mass and accounting for the
corresponding Malmquist bias \citep{mortonson11}, does not create any
tension with the reference $\Lambda$CDM model.  In order to compare
our data with the theoretical predictions, we computed the mass of
CXO1415 for a given overdensity with respect to the background density
(as opposed to the critical density, which is more commonly used)
according to the masses defined in \citet{tinker08}.  Therefore we
measure the mass of CXO1415 within an overdensity of $\Delta = 600$
with respect to the background density, finding
$M_{600}=2.0^{+0.7}_{-0.5}\times 10^{14}M_\odot$.  Finally, we find
that six clusters with the mass of CXO1415 are expected at $z \geq 1.46$
in $70$ deg$^2$.  If we consider a 20\% higher mass as expected in the
presence of a bias due to the violation of the hydrostatic equilibrium
(as discussed in Section 2.5), we find that only three clusters are
expected at $z \geq 1.46$ in $70$ deg$^2$.  Therefore, even if CXO1415
is not an extreme cluster, it is clear that a deep investigation of
the entire {\it Chandra} archive in search of faint extended sources
can provide a sizeable number of clusters with masses $\sim 10^{14}
M_\odot$.  Such a sample can provide tight constraints on the
cosmological parameters and possibly some discrepancy with respect to
the reference $\Lambda$CDM.  On the other hand, to achieve the same
accuracy on the mass on the entire sample, a large amount of {\it
  Chandra} time would be required.

\begin{figure*}
\begin{center}
\includegraphics[width=15cm,angle=0]{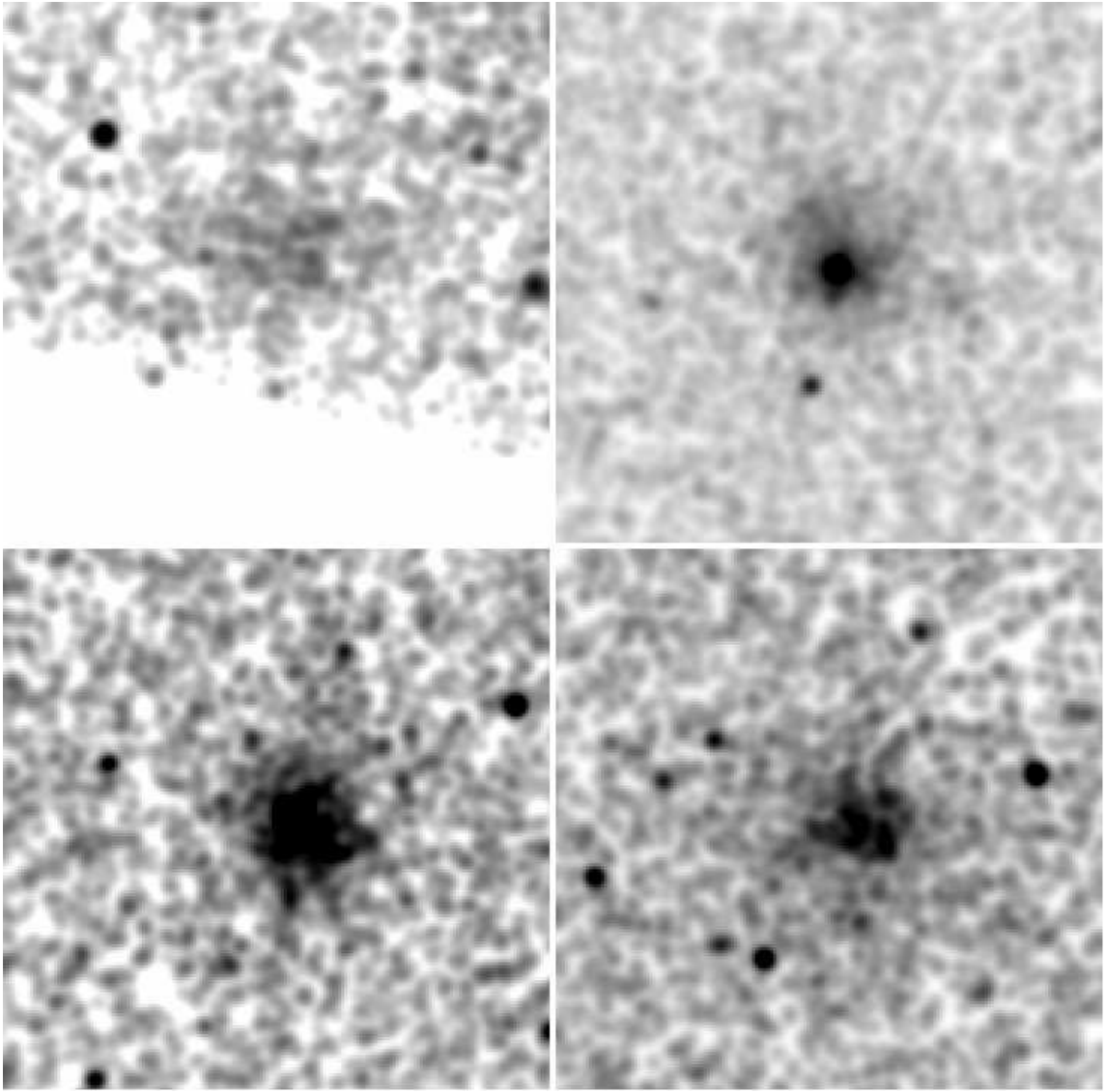}
 \caption{{\it Chandra} images of four high-redshift massive clusters
   observed with deep {\it Chandra} observations.  The four clusters are all 
   re-imaged at the same redshift of CXO1415 ($z=1.46$) and their
   observations rescaled to a 100 ks ACIS-S observation (including
   CXO1415) in order to offer an immediate comparison.  Top left:
   CXO1415; top right: WARPJ1415; bottom left: XMM2235; bottom right:
   RXJ1252.  Images are smoothed with a 1 arcsec Gaussian kernel and
   cover $3 \times 3$ arcmin.}
 \label{sb_comparison}
\end{center}
\end{figure*}

The mass determination of CXO1415 is also difficult
as it appears to be less luminous than other distant
massive galaxy clusters.  Its total\footnote{The total luminosities
  are obtained by integrating the surface brightness distribution up
  to the nominal virial radius.  Although the large majority of the
  emission is concentrated towards the center, the total luminosity is
  estimated to be about a factor $1.4$ larger than what was actually
  measured within the extraction region.} soft-band luminosity is
estimated to be $L_{X}[0.5-2.0]=(0.93 \pm 0.10) \times 10^{44}$ erg
s$^{-1}$.  Since the measured mass of WARPJ1415 at $z=1.03$ is
comparable to that of CXO1415, it is interesting to compare their
luminosities.  We find that the total luminosity of WARPJ1415 is
$L_{X}[0.5-2.0] = (4.0\pm 0.2) \times 10^{44}$ erg s$^{-1}$.  If we
remove the contribution of the cool core of WARPJ1415, using the
double beta model fitting of \citet{santos12}, we find a total
luminosity of $L_{X}[0.5-2.0]\sim (3.4 \pm 0.2) \times 10^{44}$ erg
s$^{-1}$, which is the correct value to be compared with CXO1415.
Therefore, we find that CXO1415 is more than three times less luminous 
with respect to WARP1415.  

It is also interesting to compare these values with those expected from the
empirical, redshift-dependent $M$-$L$ relation presented in
\citet{2009Vikhlinin}.  We find that WARPJ1415 is expected to have
$L_{Xexp}=(1.6\pm 0.4) \times 10^{44}$ erg s$^{-1}$, while CXO1415
$L_{Xexp}=(2.0\pm 0.5) \times 10^{44}$ erg s$^{-1}$.  The expected
values do not include an uncertainty of a factor $\sim 1.5$ due to the
observed intrinsic scatter in luminosity for a fixed mass.
Remarkably, WARPJ1415 is scattered above the average $L$-$M$ relation
by a factor of 2, while CXO1415 is scattered below it by about the same
factor.  If we consider the most massive X-ray cluster investigated
so far at high $z$, XMM2235 \citep{2009Rosati}, we find once again that
the measured total luminosity $L_{X}[0.5-2.0] = (3.0 \pm 0.15) \times
10^{44}$ is lower than the expected one by a factor of 2.  Finally,
the less massive RXJ1252 \citep{2004Rosati} at $z=1.235$ is consistent
with the average $L$-$M$ relation.

\begin{figure}
\begin{center}
\includegraphics[width=8.5cm,angle=0]{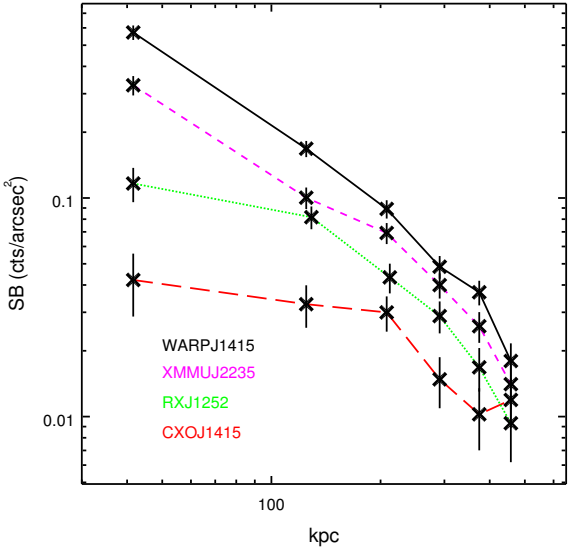}
 \caption{Radial profiles from the four clusters re-imaged at $z=1.5$
   shown in Figure \ref{sb_comparison}: WARPS1415 (solid black line);
   XMMUJ2235 (dashed magenta line); RXJ1252 (dotted green line);
   CXO1415 (long-dashed red line).  }
 \label{profiles}
\end{center}
\end{figure}

The striking difference of the appearance of the four clusters is
shown in Figure \ref{sb_comparison}, where XMM2235, WARPJ1415, and
RXJ1252 are re-imaged at the same redshift of CXO1415 for a
hypothetical 100 ks exposure with {\it Chandra} ACIS-S.  Although very small, K-correction
effects are taken into account, and the proper
background for a 100 ks ACIS-S observation is simulated.
Therefore, these images can be considered as \textit{real data}, since
they were obtained simply by trimming real exposures in order to
achieve the exact brightness that each cluster would have at $z=1.46$
for a 100 ks ACIS-S observation\footnote{We also note that the profile
  of CXO1415 shown in Figure \ref{sb_comparison} is noisier than that
  shown in this paper, since it refers only to $\sim 1/3$ (100 ks) of
  the total exposure time.}.  The only artificial interventions are
the almost negligible resizing (since they are all high-$z$ clusters
and the angular distance has a weak dependence on $z$ for $z>1$) and the
added background in order to be consistent with a real ACIS-S
observation.  Thanks to these properties, the images offer an
immediate visual comparison of the four clusters as if all of them
were at the same redshift of $z=1.46$. We perform a simple radial
profile analysis on the four images shown in Figure \ref{profiles}.
Clearly, the differences in luminosity are not only due to the
presence of a cool core, but also to the overall normalization of the
brightness profiles.  The diversity of this small sample suggests an
increasing spread in the properties of high-$z$ clusters, possibly
related to the transition from the formation of protoclusters with
little ICM emission ($z\ge 2$) and the first virialized massive
clusters around $z\sim 1.5$.  A large fraction of clusters at this
redshift may be in the first stage of collapse, an occurrence that
could explain both the low luminosity of CXO1415 and its elongated
shape.  We argue that the deep X-ray investigation of very high
redshift clusters will provide useful insights into the ICM and
dynamical properties of clusters in this crucial transition epoch.
A high-z cluster that will be observed with a deep exposure by
\textit{Chandra} in Cycle 14 is XMMUJ0044 at $z=1.578$
\citep{santos11}.  Another remarkable target is the IR-selected
cluster IDCS1426 at $z=1.75$ \citep{stanford2012}, which has a flux five
times higher than CXO1415.  Unfortunately, at present this cluster has
only a 10 ks Chandra ACIS-I exposure and therefore no X-ray mass estimate.  
We note, however, that IDCSJ1426 will be observed 
for 100 ks in {\sl Chandra} AO14 (PI M. Brodwin).

Finally, as for the chemical properties of the ICM, we notice that the
measured Fe abundance of CXO1415 is supersolar, but the 1 $\sigma$
error is large enough to make it consistent with the average values
measured at $z\sim 1.3$ \citep[see][]{2003Tozzi,2007Balestra}.  Given
the flux of CXO1415, an exposure time exceeding 1Ms would be necessary
to measure the Fe abundance with a 20\% error and to confirm the Fe
excess.  This would be extremely interesting though, since this value
cannnot be traced back to the presence of a cool core hosting a brightest cluster galaxy (BCG).
Both the low value of the $c_{SB}$ parameter and the absence of a
clear BCG in the optical image exclude the presence of a strong cool
core.  If confirmed, the high Fe abundance value would be another
striking difference with respect to low-z clusters, where supersolar
Fe abundances are always associated to cool cores and clear BCG
\citep{degrandi04,degrandi09}.  In addition, the lookback time of 9.2
Gyr can put strong constraints on the early phases of ICM enrichment
by Type Ia supernovae.  This is clearly a case for future X-ray
facilities with high effective area and high angular resolution

To summarize, although the number of cluster candidates in the range
$1<z<2$ is steadily increasing, mostly thanks to IR surveys, very few 
deep X-ray
studies are available.  Targets drawn from non-X-ray selected samples may turn
out to be very faint in X-ray, and
deep {\it Chandra} exposures will remain  the only means to physically
characterize such faint sources.  It is clear that only a sensitive,
survey-dedicated mission can provide a major breakthrough in this
field.  Even though eROSITA \citep{2010Predehl,2011Cappelluti} will finally
provide an X-ray all-sky coverage 20 years after ROSAT, it
does not have the sensitivity or the spectral response needed to find
and characterize clusters at fluxes as low as $few \times 10^{-14}$
erg s$^{-1}$ cm$^{-2}$.  On the other hand, a wide field imager with CCD
spectral resolution and an angular resolution as good as 5 arcsec
(HEW) has been shown to be highly effective in this science case 
\citep[see][]{2010WFXT,2010WFXT2}, and in many
others concerning the rich, high-z X-ray sky.

\section{Conclusions}

In this paper, we presented the deep X-ray observation of the most
distant X-ray cluster of galaxies discovered with \textit{Chandra} so
far.  CXO1415 was serendipitously discovered in the WARPJ1415 field,
and imaged in the deep exposure aimed at studying the ICM properties
of WARPJ1415 at $z=1.03$.  Even though we do not have optical
spectroscopy, our rich multi-wavelength observations allowed us to
estimate a robust photometric redshift, based on the CMR technique
and on standard SED fitting procedures.  Our photometric analysis is
further strengthened by the presence of WARPJ1415 in the imaging
fields, which acted as redshift calibrator.  The photometric redshift
of CXO1415 turns out to be $z_{phot}=1.52 \pm 0.06$.  This result is
confirmed by the detection of the $K_\alpha$ Fe emission line in the
X-ray spectrum, which gives $z_X = 1.46\pm 0.025$.  Contrary to the few
currently known very high-redshift clusters, we are able to obtain a
robust measure of the ICM temperature and total mass within 300 kpc
and a reliable extrapolation up to $R_{500}$.  Here we summarize our
main results:

\begin{itemize}

\item CXO1415 is detected with a $S/N\sim 11$ within a circle with a
  radius of 24 arcsec.  The surface brightness can be fitted by a
  beta model; the soft-band flux within $r=24\arcsec$ is equal to
  $S_{0.5-2.0 keV}=(6.5 \pm 0.3) \times 10^{-15}$ erg s$^{-1}$
  cm$^{-2}$;

\item we identify the Fe $K_\alpha$ complex line with a $\sim 2.8
  \sigma$ confidence level, showing that the measure of redshift
  through X-ray spectral analysis is possible up to the highest
  redshift where X-ray clusters are currently detected;

\item the spectral fit with a {\tt mekal} model gives
  $kT=5.8^{+1.2}_{-1.0}$ keV and a Fe abundance $Z_{Fe} =
  1.3_{-0.5}^{+0.8}Z_\odot$;

\item the total mass measured at $R_{2500}=(220\pm 55)$ kpc is $M_{2500} =
  8.6_{-1.7}^{+2.1} \times 10^{13} M_\odot$.  The total mass is
  measured up to 300 kpc, where it amounts to $M(r<300 {\tt kpc}) =
  1.38_{-0.28}^{+0.33} \times 10 ^{14}M_\odot$, while the ICM mass is
  $M_{ICM} (r<300 {\tt kpc}) = 1.09_{-0.20}^{+0.30}\times 10^{13}M_\odot$,
  resulting in a ICM mass fraction of $\sim 10$\%;

\item by fixing the slope of the density profile to the value measured
  at 300 kpc and assuming a mildly decreasing temperature profile as
  observed in local clusters, we are able extrapolate the mass
  measurement up to $R_{500} = 510_{-50}^{+55}$ kpc, finding $M_{500} =
  2.07_{-0.50}^{+0.70} \times 10 ^{14}M_\odot$; this value is
  consistent with those obtained using empirical relations between
  X-ray observable and total mass;

\item considering its mass, CXO1415 appears to be underluminous,
  particularly if compared with other high-z massive clusters;

\item the color-magnitude diagram of cluster member candidates shows
  both a red sequence and a significant fraction of blue,
  irregular-morphology galaxies.  A spectroscopic follow up is needed
  to perform an in-depth study of the galaxy population, which is
  expected to show the signature of the red sequence assembly around
  $z\sim 1.5$;

\item when compared with the expectations for a $\Lambda$CDM universe
  based on the mass function of \citet{tinker08}, CXO1415 appears to
  be a typical cluster at $z\sim 1.5$ for a WMAP cosmology
  \citep{komatsu11}; however, the redshift and the total mass of
  CX01415 place it among the sample of massive, distant galaxy clusters
  which may be used to test the standard $\Lambda$CDM model once a
  complete census of the {\it Chandra} archive is performed.

\end{itemize}

We note that the realm of high-z clusters ($z\sim 1.5$ and higher,
when the lookback time is larger than 9 Gyr) is at the limit of the
capability of \textit{Chandra} and well beyond what XMM-Newton and eROSITA
can reach. At present, only an handful of sources have the X-ray
coverage sufficient for a detailed analysis.  While the {\it Chandra} and
XMM-Newton missions will somewhat increase this sample, it is
not realistic to foresee a significant increase of data on massive
high-z clusters with the required high S/N.  In the near future, this
field is likely to be dominated by SZ observations and other near-IR
large-area surveys, which will unveil more and more clusters at $z\gg
1$.  However, X-ray observations will still be required for a physical
characterization of these systems and a vast range of astrophysical
and cosmological applications. Only a wide-field, high angular
resolution X-ray mission with a large collecting area and good 
spectral resolution up to 7 keV seems to be able to match such
requirements.

\acknowledgements

The authors thank Dunja Fabjan and Renee Fassbender for discussion.
We also thank the referee, Manolis Plionis, for useful comments.
We acknowledge financial contribution from contract ASI-INAF
I/088/06/0 and ASI-INAF I/009/10/0, and from the PD51 INFN grant.
This work was carried out with \textit{Chandra} Observation Award
Number 12800510 in GO 12.  PR  acknowledges partial support
from the DFG cluster of excellence Origin and Structure of the
Universe program.  Results are also based on data collected at
Subaru Telescope, Okayama Astrophysical Observatory, Kiso
Observatory (University of Tokyo) and obtained from the SMOKA, which
is operated by the Astronomy Data Center, National Astronomical
Observatory of Japan.  This research has made use of the NASA/IPAC
Infrared Science Archive, which is operated by the Jet Propulsion
Laboratory, California Institute of Technology, under contract with
the National Aeronautics and Space Administration.

\bibliography{references_CXO1415}

\end{document}